\documentclass[preprint,nofootinbib,aps,superscriptaddress,eqsecnum]{revtex4-1}
 \pdfoutput=1
\usepackage{amsmath,amssymb,graphicx,subfigure}
\usepackage{color}
\usepackage{hyperref}

\newcommand{\beq}{\begin{equation}}
\newcommand{\eeq}{\end{equation}}
\newcommand{\bea}{\begin{eqnarray}}
\newcommand{\eea}{\end{eqnarray}}

\def\nn{\nonumber}

\def\ra{\rightarrow}

\begin{document}
% \begin{center}
\title{Neutrino mass and the Higgs portal dark matter in the ESSFSM}
%\vspace {1.0cm}

\author{Najimuddin Khan}\email{najimuddin@prl.res.in, khanphysics.123@gmail.com} 
\affiliation{ Theoretical Physics Division, 
Physical Research Laboratory, Ahmedabad - 380009, India}
\affiliation{
Centre for High Energy Physics, Indian Institute of Science, Bangalore - 560012, India\vspace{1.80cm}}

\begin{abstract}
We extend the standard model with three right-handed singlet neutrinos and a real singlet scalar. We impose two $Z_2$ and $Z_2^\prime$-symmetries.  
We explain the tiny neutrino mass-squared differences with two $Z_2$- and $Z_2^\prime$-even right-handed neutrinos using the type-I seesaw mechanism. 
The $Z_2$-odd fermion and the $Z_2^\prime$-odd scalar can both serve as viable dark matter candidates.
We identify new regions in the parameter space which are consistent with relic density of the dark matter, recent direct search experiment LUX-2016, XENON1T-2017 and LHC data. 
\end{abstract}

\keywords{Neutrino Mass; Dark Matter}

\renewcommand*{\thefootnote}{\fnsymbol{footnote}}
\maketitle
%\newpage

\section{Introduction}

The found Higgs boson at the Large Hadron Collider (LHC)~\cite{Aad:2012tfa,Giardino:2013bma,Chatrchyan:2012xdj}, completes the search for the particle content of the standard model (SM). 
The hierarchy problem related to the Higgs boson mass has motivated a plethora of models such as supersymmetry, extra dimensions, etc. in which the fine-tuning is reconsidered. 
However, an inevitable consequence of these models is that the new physics should lie close to the TeV scale. 
Non-observations~\cite{Ruhr:2016xsg} of any new physics from the collider experiments imply that the Higgs hierarchy issue is reverting back to being an unsolved open problem.

In addition  the SM is unable to explain some physical phenomena in the Nature such as the existence of massive neutrinos, the presence of dark matter (DM), the observed matter-antimatter asymmetry, etc.
In the SM, by construction, the
neutrinos are massless as it does not include right-handed neutrinos. However from the neutrino
oscillation experiments, we got convinced that at least two neutrinos have non-zero mass.
The neutrino oscillation experiments have given information about the mass squared differences between neutrino mass eigenstates. However the individual value of the masses is not
yet known. It has been seen that the sum of the three
neutrino masses is less than $\sim$0.1 eV~\cite{Ade:2015xua,Giusarma:2016phn,Vagnozzi:2017ovm} which is consistent with the cosmological measurements. Individual masses and the basic nature of neutrinos,
i.e., whether they are Dirac or Majorana particles are still an open question.

As neutrino masses are very tiny compared to the other fermion masses, it is believed that the
mechanism behind neutrino mass generation is different from the other fermions. The other fermions are obtained mass through the Higgs mechanism. The most popular natural explanation of small neutrino masses is the see-saw mechanism. There are broadly three classes of such models
namely type-I, type-II, and type-III see-saw models requiring involvement of right-handed
neutrinos, a $SU(2)_L$ triplet scalar with hypercharge $Y=2$ and $SU(2)_L$ hyperchargeless triplet fermions respectively.
The minimal scenario in this respect is the canonical
type-I seesaw mechanism, in which the SM is extended by a right-handed Majorana neutrinos~\cite{Minkowski:1977sc,seesaw1,seesaw2,Mohapatra:1979ia,Mitra:2011qr,Karmakar:2014dva,Nath:2016mts}. The TeV-scale seesaw mechanism has been discussed in Refs.~\cite{Boucenna:2014zba,Deppisch:2015qwa,Khan:2012zw}. Including extra scalar fields, it has been studied in Refs.~\cite{Chakrabortty:2012mh,Ghosh:2017fmr,Garg:2017iva,Bhattacharya:2016qsg}.

Various kinds of astrophysical observations such as anomalies in the galactic rotation curves, gravitational lensing effects in the Bullet cluster, excess gamma rays\footnote{The excess gamma rays from the galactic centers may come from other sources like pulsars.} from the galactic centers, etc., have indicated the existence of DM in the Universe. The cosmological measurements of tiny anisotropies in Cosmic Microwave Background Radiation (CMBR) by the WMAP and Planck Collaboration~\cite{Ade:2015xua} suggest that the Universe made of $69\%$ dark energy, $27\%$ dark matter and $4\%$ ordinary matter.

Astrophysical and cosmological data can tell us about the total amount/density of the DM of the Universe. There is still no consensus on what it is composed of and the properties are still unknown.
A study for the possibilities of different kinds of baryonic or non-baryonic DM candidates have been discussed in Ref.~\cite{Khan:2017xyh}. The weakly interacting massive particles (WIMPs) are the best viable DM candidates.
No evidence of the WIMP has been found from the direct detection experiments such as XENON100~\cite{Aprile:2012nq}, LUX~\cite{Akerib:2013tjd,Akerib:2016vxi}, XENON1T~\cite{Aprile:2017iyp}, etc. As these DM-nucleon scattering experiments still have not found any signature in the detector, these experiments have ruled out low mass ($10-50$ GeV) regions in the parameter space of a $Z$- and Higgs $h$-portal DM. Recent LUX-2016~\cite{Akerib:2016vxi} data has also excluded the mass range $65-550$ GeV
of a $h$-portal fermionic~~\footnote{It depends on the mixing angle between Higgs and singlet scalar} DM model~\cite{Ettefaghi:2017vbh} and scalar DM models~\cite{Ettefaghi:2017vbh,Athron:2017kgt,Khan:2014kba}.
It indicates that we may need the multi-component DM particles to explain the experimental data. We may detect these DMs in the more efficient detector in the future experiments. 
Multi-component DM model is needed~\cite{DuttaBanik:2016jzv} to explain the Galactic
Center gamma ray excess~\cite{Fermi-LAT:2016uux} and the colliding galaxy cluster~\cite{Harvey:2015hha,Kahlhoefer:2015vua,Campbell:2015fra} simultaneously.
Multi-component DM models have been considered in Refs.~\cite{Hannestad:2010yi,Bae:2013hma} in various models which also includes neutrino, Axion, supersymmetric particles.
Various models with two WIMP candidates could lead to typical signatures at different mass scale, have been studied in Refs~\cite{Ma:2006km,Zurek:2008qg,Batell:2010bp,Fukuoka:2010kx,Belanger:2012vp,Aoki:2012ub,Ivanov:2012hc,Chialva:2012rq,Heeck:2012bz,Modak:2013jya,Aoki:2013gzs,Geng:2013nda,Kajiyama:2013rla,Bhattacharya:2013hva,Biswas:2013nn,Biswas:2015sva,Bian:2014cja,Bhattacharya:2013asa,Bian:2013wna,Gu:2013iy,DiFranzo:2016uzc,Bhattacharya:2016ysw}.

We add three right-handed $SU(2)$ singlet fermions and a singlet scalar to the SM. We also impose two $Z_2$ and $Z_2^\prime$ symmetry.
All SM and the first two fermion fields are even under these $Z_2$ and $Z_2^\prime$ transformations. The
Dirac mass terms can be formed using these fermions and the SM neutrinos. We use the type-I seesaw mechanism to explain the tiny neutrino mass-squared differences and the mixing angles which are observed by the neutrino oscillation experiments.
The third $Z_2$-odd fermion and $Z_2^\prime$-odd scalar both can serve as viable DM particles in this work. Moreover, the requisite rate of annihilation is ensured by postulating some $Z_2$ and $Z_2^\prime$ preserving dimension four and five operators for the scalar and fermion particles respectively. The four-point interaction term of the extra fermions and scalar can be obtained from other five-dimension operators~\cite{Chaudhuri:2015pna}. The interaction term of the third fermion and the scalar allows a larger region of the parameter space than what we would have had with a single DM particle (either fermion or scalar) alone.
This interplay brings an enriched DM phenomenology compared to the other models having fermion or scalar DM particle.
The region of DM masses $65-550$ GeV of a fermionic or scalar Higgs portal DM model is excluded from the present LUX experimental data. In this model, we show that the region with masses $50-550$ GeV up to 300 TeV is still allowed by the direct search experiments. Hence, we feel a desirable feature of our model for future study.

The plan of the paper is as follows. In section~\ref{S:theory}, we present the theoretical framework of our extended singlet scalar fermionic standard model (ESSFSM). We also discuss the
diagonalization procedure to get the neutrino mass matrix and the relic density calculation of two dark matter particles. We show the detailed constraints on this model in section~\ref{sec:constraints}. We present our numerical results and show the allowed region in the parameter spaces from the neutrino mass and mixing angle, relic density and direct detection in section~\ref{sec:result}. Finally, we conclude in section~\ref{sec:conclusion}.

\section{Theoretical framework of the model}
\label{S:theory}

In this section, we give a description of our model. We add three right-handed neutrinos and a scalar to the SM Lagrangian. These extra particles are singlet under $SU(2)$ transformation.
We impose two $Z_2$ and $Z_2^\prime$ symmetry such that the SM fields and first two right-handed neutrinos are even under these $Z_2$ and $Z_2^\prime$ transformations. The third right-handed neutrino is odd (even) under $Z_2$ ($Z_2^\prime$) transformation whereas the scalar field is even (odd) under $Z_2$ ($Z_2^\prime$) transformation.
 The $Z_2 \times Z_2^{\prime}$ quantum numbers of the SM fields and extra right-handed neutrinos and scalar fields are summarized in Table~\ref{tabZ2}.
%%%%%%%%%%%%%%%%%%%%%%%%%%%%%%%%%%%%%%%%%%%%%
\begin{table}[h!]
\begin{center}\scalebox{0.7}{
\begin{tabular}{|c||c|c|c|}
\hline
~~~~~~Fields~~~~~~ &\multicolumn{3}{c|}{ charged under $SU(2) \times Z_2 \times Z_2^{\prime}$ transformation}\\
\cline{2-4} 
 & ~~~~~~$SU(2)$ ~~~~~~& ~~~~~~~~~$Z_2$~~~~~~~~~& ~~~$Z_2^{\prime}$ ~~~ \\
\hline
$Q_L = \begin{pmatrix}
u_L\\d_L
\end{pmatrix} $&    $2$&$1$&$1$\\

$u_R,d_R$~&   $1$&$1$&$1$\\
$L = \begin{pmatrix}
\nu_l\\l^{-}
\end{pmatrix} $&    $2$&$1$&$1$\\

$l_R$~&   $1$&$1$&$1$\\

$\Phi = \begin{pmatrix}
G^+\\\frac{h+v+i G^0}{\sqrt{2}}
\end{pmatrix} $&    $2$&$1$&$1$\\

$\nu_{s,1}\,, \,\nu_{s,2}$~&    $1$&$1$&$1$\\
$\nu_{s,3}$~&    $1$&$-1$&$1$\\
$S$~&    $1$&$1$&$-1$\\
\hline
\end{tabular}}
\end{center}
\caption{ 
The $Z_2 \times Z_2^{\prime}$ quantum numbers. $u$ represents the $up$-$type$ quarks of the three generations $u, ~c, ~t$ and $d$ stand
for the $down$-$type$ quarks $ d, ~s, ~b$. The charged leptons are denoted by $l = e, \mu, \tau$ with
the corresponding left-handed neutrinos $\nu_l$. $\Phi$ is the SM Higgs doublet. $G^+ ~(G^0)$ stand for the charged (neutral) Goldstone boson. $L,R$ stand for left- and right-handed chirality of fermions.}
\label{tabZ2}
\end{table}
%%%%%%%%%%%%%%%%%%%%%%%%%%%%%%%%%%%%%%%%%%%%%%%%%%%%%%%%
The $Z_2\times Z^\prime_2$-even neutrinos are free to mix with the usual SM neutrinos and therefore generate the neutrino masses through the type-I seesaw mechanism. 
These symmetries are prohibited the coupling of 
an $odd$ number of the third fermion and/or the scalar particle to the SM particles.
The part of Lagrangian that invariant under $SU(2)\times U(1) \times Z_2 \times Z_2^{\prime}$ transformation is given by
\bea
\mathcal{L} =  \frac{i}{2} \overline{\nu}_{s,a} \, \slash \hspace{-2.5mm} \partial \nu_{s,a}
-\frac{1}{2} \overline{M}_{\nu_{s,a}}\overline{\nu}_{s,a} \nu_{s,a}^{c}  + \frac{1}{2} \partial_\mu S \partial^\mu S - \frac{\mu_S}{2} S^2 -\frac{\lambda_S}{4!} S^4 \,,
\label{eq:Lag1}
\eea
where summation over $a$ is implied, with
$a = 1,2,3$ denote generation indices for the right-handed fermions. $c$ stands for the charge conjugation. 
The mutual interaction terms of the SM Higgs, left-handed leptons, the extra scalar and fermions are given by 
\bea
\mathcal{L}_{mix} &=& -  Y_{\nu,ab}\, \overline{L}_a {\Phi}^c \nu_{s,b} - \overline{M}_{\nu_s,{mn}} \, \overline{\nu}_{s,m}\nu_{s,n}^{c}   \,  -\frac{\kappa}{2} |\Phi|^2 S^2  + \frac{C_{h,mn}}{\Lambda_{h,mn}} |\Phi|^2\overline{\nu}_{s,m}\,\nu_{s,n}^{c}  \nn\\&& + \frac{C_{h,a}}{\Lambda_{h,a}} |\Phi|^2\overline{\nu}_{s,a}\,\nu_{s,a}^{c}  +  \frac{C_{S,mn}}{\Lambda_{S,mn}} S^2\overline{\nu}_{s,m}\nu_{s,n}^{c}\,   + \frac{C_{S,a}}{\Lambda_{S,a}} S^2\overline{\nu}_{s,a} \nu_{s,a}^{c} + \textrm{h.c.}
\label{eq:Lag2}
\eea
$\Phi$ is the SM Higgs doublet, $\Phi \equiv (G^+,~ (v+h+i G^0)/\sqrt{2})^T$, where the
$G^\pm$ and $G^0$ are the Goldstone bosons and $h$ is the SM Higgs. ${\Phi}^c$ stands for charge conjugate of ${\Phi}$. $L\equiv (\nu_l,~ l)^T$ with $l=e,\mu$ and $\tau$ are the left-handed lepton doublet. $b=1,2$ does not assume the third index as the third fermion is odd under $Z_2$-symmetry. The indices $m \neq n =1,2$; hence the second term in eqn~\ref{eq:Lag2} generates the mixing mass term between two $Z_2$- and $Z_2'$-even neutrinos.
After electroweak (EW) symmetry breaking, the fourth term, i.e., the dimension-five operator also gives an additional mixing mass term.
The Higgs to extra neutrinos couplings are also generated from the dimension-five operators (fourth and fifth term in eqn~\ref{eq:Lag2}). 
This will lead to the Higgs boson decay into these extra neutrons. As the $Z_2$- and $Z_2'$-even neutrinos are considered to be very heavy, the partial decay width of the Higgs to these neutrinos is zero.
As we are allowing these dimension-five operators in the Lagrangian, for completeness we also add the other dimension-five operators $\frac{C_{S,mn}}{\Lambda_{S,mn}} S^2\overline{\nu}_{s,m}\nu_{s,n}^{c}\,$ and $\frac{C_{S,a}}{\Lambda_{S,a}} S^2\overline{\nu}_{s,a} \nu_{s,a}^{c}\,$ as well, which in turn give more room in the parameter space to maneuver.
In this work, we focus on the dominant dimension-five operators related to the neutrino and Higgs
portal dark matter physics, i.e., those involving at least one Higgs and neglect other possible operators which are allowed by the SM gauge and $Z_2 \times Z_2^\prime$ symmetries.
$\Lambda$'s are the cut-off scales for the new physics. In our calculation, we assume $\Lambda_{h,a}=\Lambda_{S,a} = \Lambda_{h,mn}=\Lambda_{S,mn}\equiv \Lambda$. 
$C_{h,a}$, $C_{S,a}$, $C_{h,mn}$ and $C_{S,mn}$ are dimensionless coupling parameters.
The cut-off scale $\Lambda$ and $C_{h,12}$ and the Yukawa couplings $Y_{\nu,ab}$ are important to explain the neutrino oscillation observables. Whereas $\Lambda$, $C_{h,3}$, $C_{S,3}$ and $\kappa$ could change the masses and coupling strength of DM particles to the Higgs. In addition these could alter the self-annihilation interaction probability of the heavier DM particles into the lighter DM particles. Hence, these parameters play a crucial role to calculate the relic density of the DM particles ${\nu_{s,3}}$ and $S$.
The masses of the DM particles are given by
\beq
	M_{\nu_{s,3}} = \overline{ M}_{\nu_{s,3}} - \frac{C_{h,3}}{\Lambda} v^2\qquad \text{and} \qquad M_{S}^2 = \mu_S^2 + \frac{1}{2} \kappa v^2,
\eeq
and the coupling strength of the DM candidates with the Higgs can be written as
\bea
h \overline{\nu}_{s,3}\nu_{s,3} ~~:   \qquad \frac{C_{h,3}}{\Lambda} \qquad \qquad \text{and} \qquad \qquad
h SS  ~~: \qquad   \frac{\kappa}{2} v.
\eea

The parameter ${C_{S,3}}$ is responsible for the annihilation of the $SS \leftrightarrow \overline{\nu}_{s,3}\nu_{s,3}$. This process reduces the number density of the heavier DM till the freeze-out.

It is also important to note that the gauge boson $B_\mu$ and/or $ W_\mu^i$ interactions terms are not present in the kinetic part of the Lagrangian (see eqn.~\ref{eq:Lag1}). Therefore, this model does not have any extra gauge-boson contribution to the DM-nucleus scattering cross-section which is allowing larger region in the parameter space from the direct detection experiments. This is the specialty of the presence of real singlet scalar and fermion in the ESSFSM.

\subsection{Diagonalisation procedure of the type-I seesaw matrix and non-unitary of PMNS matrix}
Here, we show the diagonalization procedure~\cite{Casas:2001sr,Bambhaniya:2016rbb} of type-I seesaw mechanism to generate tiny neutrino mass-squared difference~\cite{Capozzi:2016rtj,Esteban:2016qun}.
In this model, $5 \, \times 5$ neutrino mass matrix in the basis $(\nu_l, \nu_s)$ can be written as
\beq
M_\nu \, = \, \begin{pmatrix}
 0 & {M}_D \\
 {M}_D^T &  {M}_{\nu_s}
\end{pmatrix} \label{matmnu},
\eeq
where, the Dirac mass ${M}_D$ and Majorana mass ${M}_{\nu_s}$ terms can be written as
  
\beq
{M}_D = \begin{pmatrix} 
 Y_{\nu,{11}} \, v &  Y_{\nu,{12}} \, v\\
  Y_{\nu,{21}} \, v &  Y_{\nu,{22}} \, v\\
   Y_{\nu,{31}} \, v &  Y_{\nu,{32}} \, v
 \end{pmatrix}\qquad {\rm and} \qquad {M}_{\nu_s} =  \begin{pmatrix}
  M_{11} & M_{12}\\
M_{12} & M_{22}
\end{pmatrix}
\label{MasMatrixs}.
\eeq  
Here, $M_{11} = \overline{ M}_{\nu_s,1} - \frac{C_{h,1}}{\Lambda} v^2$, $M_{22}= \overline{ M}_{\nu_s,2} - \frac{C_{h,2}}{\Lambda} v^2$ and $M_{12} =\overline{ M}_{\nu_s,12} - \frac{C_{h,12}}{\Lambda} v^2$

Using a $5 \times 5$ unitary matrices~\cite{Xing:2005kh,Grimus:2000vj}, one can diagonalize the neutrino mass matrix $M_\nu$ in eqn~\ref{matmnu}. It is given by
\beq U^T \,M_\nu \, U \,\, = \,\, M_\nu^{diag}  \eeq
 where, $M_\nu^{diag} \, = \,\textrm{ diag}\,(m_i,\, M_j )\,$ with mass eigenvalues $\,m_i \, (i =
1,\, 2,\, 3)\,$ for three light neutrinos and $\,M_j\, (j = 1, 2)\,$ for two heavy neutrinos respectively. In this calculation, we have two non-zero mass eigenstates of light neutrinos. We consider $m_1$ to be zero. 
In the limit ${M}_D^2 << {M}_{\nu_s}^2$, the matrix $U$ can be expressed as~\cite{Bambhaniya:2016rbb},
\beq
 U =  W\,T =\, \begin{pmatrix}
 U_L & V \\
S & U_H
\end{pmatrix} \,=\, 
\begin{pmatrix}
 (1-\frac{1}{2}\epsilon)U_\nu & {M}_D^* ({M}_{\nu_s}^{-1})^*U_R \\
-{M}_{\nu_s}^{-1} {M}_D^T \, U_\nu & (1-\frac{1}{2}\epsilon ')U_R
\end{pmatrix} \label{UL2},
\eeq
where, $\, U_L,\, V,\,S\,\, \textrm{and} \,\,U_H\,$ are $3 \times 3 \,,\, 2 \times 3 \,,\, 3\times 2 \,\, \, \textrm{and}\,\, 2\times 2\, $ 
matrices respectively which are not unitary.
The unitary $W$ matrix which brings the full $5 \times 5$ neutrino matrix in the
block diagonal form as
\beq W^T \begin{pmatrix}
 0 & \hat{M}_D \\
 {M}_D^T & {M}_{\nu_s}
\end{pmatrix} W \,=\,   \begin{pmatrix}
 m_{light} & 0 \\
 0 & M_{heavy}
\end{pmatrix}     \eeq
Another unitary matrix $T \, = \, \textrm{diag}\,(U_\nu, U_R )$ matrix again diagonalizes the mass matrices in the light and heavy sectors are
appearing in the upper and lower block of the block diagonal matrix respectively. In the above-stated limit, one can then write the light neutrino mass matrix to the leading order as
\beq  
 m_{light}\,=\,M_D {M}_{\nu_s}^{-1} M_D^T
\eeq 
%%%%%%%%%%%%%%%%%%%%%%%%%%%%%%%%%
In eqn.~\ref{UL2}, $U_L$ corresponds to the PMNS matrix which acquires a non-unitary correction $(1-\frac{\epsilon}{2})$ due to the presence of heavy neutrinos. The characterizations of non-unitarity are denoted by the notations $\epsilon$ and $\epsilon'$. These are given by~\cite{Casas:2001sr}
%%%%%%%%%%%%%%%%%%
\beq
 \epsilon \, = \, {M}_D^* {M}_{\nu_s}^{-1*} {M}_{\nu_s}^{-1} {M}_D^T \qquad {\rm and}\qquad \epsilon ' \, = \, {M}_{\nu_s}^{-1} {M}_D^T  {M}_D^* {M}_{\nu_s}^{-1*}
\eeq
%%%%%%%%%%%%%%%%%%

\subsection{Relic density calculation of the two-component dark matter}
In order to calculate the relic abundance of two-component DM in the present formalism, we need to solve the relevant coupled Boltzmann eqns.~\cite{Belanger:2014vza}  
\bea
\frac{dn_{\nu_{s,3}}}{dt} + 3Hn_{\nu_{s,3}} &=& -\langle\sigma v\rangle_{{\nu_{s,3}}{\nu_{s,3}}\rightarrow XX}
(n_{\nu_{s,3}}^2-n_{{\nu_{s,3}}\,{\rm eq}}^2) \nn\\&&- \langle\sigma v\rangle_{{\nu_{s,3}}{\nu_{s,3}}\rightarrow SS}
\left(n_{\nu_{s,3}}^2-\frac{n_{{\nu_{s,3}}\,{\rm eq}}^2}{n_{S_{\rm eq}}^2}n_{S}^2\right)
\label{be1}\\
%\end{equation}
%\begin{equation}
\frac{dn_{S}}{dt} + 3Hn_{S} &=& -\langle\sigma v\rangle_{SS\rightarrow XX}
(n_{S}^2-n_{S_{\rm eq}}^2) \nn\\&&- \langle\sigma v\rangle_{SS\rightarrow {\nu_{s,3}}{\nu_{s,3}}}
\left(n_{S}^2-\frac{n_{S_{\rm eq}}^2}{n_{{\nu_{s,3}}\,{\rm eq}}^2}n_{\nu_{s,3}}^2\right)\,,
\label{be2}
\eea
where, $Z_2$-even (SM, $\nu_{s,2}$ and $\nu_{s,2}$) particles are denoted by $X$. In addition  the heavier $X$ can decay into lighter particles. $\langle\sigma v\rangle$ is the average effective annihilation cross-sections of the DM candidates which include all $n\geq2$-body final state particles. The first term on the right-hand side of eqn.~\ref{be1} and \ref{be2} indicate the contribution of annihilation to SM particles whereas the second term in both the equations take care of the contribution of the self-scattering of DM particles.
The contributions from the processes ${\nu_{s,3}} S \rightarrow {\nu_{s,3}} S$ is zero as it does not alter the number density. In the very early Universe, both of the DM candidates are in thermal and chemical equilibrium. In the non-relativistic case, if the temperature $T$ of the Universe is less than the DM masses, then the equilibrium number density takes the form
$n_{DM \, eq}= \left(\frac{M_{DM} T}{2 \pi}\right)^{3/2} exp\left(-\frac{M_{DM}}{T}\right)$.
As the temperature was falling down, some species are decoupled and contributing to the relic density. The heavier DM candidate particle decouples earlier than the lighter one. In the present Universe, they both were frozen out and giving a partial contribution in the total relic abundance $\Omega_{\rm tot}$. If the individual contributions of the fermion and scalar are $\Omega_{{\nu_{s,3}}}$ and $\Omega_{S}$, then the total relic abundance $\Omega_{\rm tot}$ can be written as
\beq
\Omega_{\rm DM} = \Omega_{{\nu_{s,3}}} + \Omega_{S} \,,
\label{omega}
\eeq
where, $\Omega_{{\nu_{s,3}}}  = \frac{M_{\nu_{s,3}}}{\rho_c}  n_{{\nu_{s,3}}}(T_0)$ and $\Omega_{{\nu_{s,3}}}  = \frac{M_S}{\rho_c}  n_S(T_0)$. $\rho_c\sim {\rm 1.05\times10^{-5}} h^2 $
${\rm ~GeV cm^{-3}}$ stands for the critical density of the present Universe, $h=0.72$ is the Hubble parameter. $n(T_0)$ is the number density of the DM at temperature $T_0$ today.

One can note that if the masses of the DM particles are degenerate, then the Boltzmann eqns.~\ref{be1} and \ref{be2} become decoupled, i.e., self-scattering cross-sections of the process $\nu_{s,3}\nu_{s,3} \leftrightarrow S S$ is very small compared to the self-annihilation cross-section of the DM. These equations describe the evolution of each DM independently. In our calculation, we use the {\tt micrOMEGAs}~\cite{Belanger:2014vza} and solve the above coupled Boltzmann equations to calculate the individual number density of the DM particles in the present Universe.

\section{Constraints on the model}
\label{sec:constraints}
The parameter spaces of this model are constrained from various theoretical considerations like  absolute vacuum stability, perturbativity and unitarity of the scattering matrix.
The absolute stability of the Higgs potential demands that the scalar potential should not approach to negative infinity along any direction of the field space at large field values. The required conditions are: $\lambda > 0$, $\lambda_S>0$ and $\kappa> -\sqrt{2 \lambda \lambda_s}/\sqrt{2}$, where $\lambda$ is the Higgs quartic coupling~\cite{Khan:2014kba}. 
Lagrangian of our model remains perturbative~\cite{Lee:1977eg,Cynolter:2004cq} for
$|\lambda| \, 
 \lesssim \, \frac{4\pi}{3}, \,\,\ |\kappa|\, \lesssim\,8\pi, \,\,\,|\lambda_{S}(\Lambda)|\, \lesssim \, 8\pi, \,\,\ C_{h,a}\lesssim \, 8\pi, \,\,{\rm and}\,\,\ C_{S,a} \lesssim \, 8\pi$. 
The parameters of the scalar part of Lagrangian (see eqns.~\ref{eq:Lag1} and \ref{eq:Lag2}) of this model are constrained by the unitarity of the scattering matrix (S-matrix). One can obtain the S-matrix by using various scalar-scalar, gauge boson-gauge boson, and scalar-gauge boson scattering amplitudes. We use the equivalence theorem~\cite{equivalence1,equivalence2,equivalence3} to reproduce the S-matrix for this model~\cite{Cynolter:2004cq}. The unitary bounds demand that the eigenvalues of this matrix should be less than $8\pi$ which imply
$
\lambda \leq 8 \pi  \quad {\rm and} \quad | 12 {\lambda}+{\lambda_S} \pm \sqrt{16 \kappa^2+(-12 {\lambda}+{\lambda_S})^2}| \leq 32 \pi
$.
%%%%%%%%%%%%%%%%%%%%%

The observed neutrino mass-squared differences and mixing angles by the neutrino oscillation experiments put stringent constraints on the parameter space of this model. The Higgs signal strength and the decay width measured by the LHC, the relic density and direct-indirect searches of DM all alone restrict the allowed parameter space considerably. We discuss these in the following.

\subsection{Bounds from the neutral fermion mass and mixing angles}
The global analysis of neutrino oscillation measurements provide the neutrino oscillation parameters for both normal and inverted hierarchies scenario. These can be found in Refs.~~\cite{Capozzi:2016rtj,Esteban:2016qun}.
The measurements of the electroweak precision observables along with other experimental data put severe constraints on the light neutrino mixing matrix $U_L$. The detailed analysis has been given in Refs~\cite{Antusch:2014woa, Antusch:2016brq}.

The L3 collaboration at the LEP had analyzed the decay channels $N \, \rightarrow \,\, e^\pm\,W^\mp$ to find the evidence of the heavy neutrino.
No signature had been found for the mass range in between 80 GeV ( with $|V_{\alpha i} |^2 \, 
\leq \, 2 \times 10^{-5} $) and
205 GeV ( with $|V_{\alpha i} |^2 \, \leq \,1$) \cite{Achard:2001qv}.
$V$ is the light-heavy mixing matrix, given in eqn.~\ref{UL2}.
This puts a lower bound on the mass of the heavy neutrino and the mixing matrix elements $V_{\alpha i}$.
$|V_{\alpha i} |^2 \,\gtrsim \, 10^{-5}$ and $3<M_{1,2}<M_Z$ region have also been ruled out from the invisible decay width of the $Z$-boson~\cite{Adriani:1992pq,Abreu:1996pa,L3}.

The experimental data~\cite{TheMEG:2016wtm} $\textrm{Br}(\mu \, \rightarrow e \, \gamma) < 4.2\times 10^{-13}$ of the flavor changing decay processes has restricts the arbitrary Yukawa coupling $Y_\nu$
%%%%%%%%
In this model, the branching ratio can be written as~\cite{Tommasini:1995ii}
\beq
\textrm{Br}(\mu \, \rightarrow e \, \gamma)  \,\, = \,\, \frac{3\alpha}{8\pi}\, |V_{ei}\,V_{i\mu}^\dagger \, f(x)|^2
\eeq
where, $ \,\, x \,=\, (\frac{M_i^2}{M_W^2})\, $, ${i=1,2}$ stands for the mass of heavy neutrinos and $f(x)$ is the slowly varying function can be found in Ref.~\cite{Tommasini:1995ii}.

\subsection{Bounds from the Higgs signal strength at the LHC}
The dominant contribution of the Higgs $h$-production cross-section is coming through the gluon fusion. In this work, the Higgs to diphoton signal strength $\mu_{\gamma\gamma}$ can be written as
\beq
\mu_{\gamma\gamma} \simeq \frac{\sigma(gg\ra h\ra\gamma\gamma)_{ESSFSM}}{\sigma(gg\ra h\ra\gamma\gamma)_{\rm SM}}= \frac{\sigma(gg\ra h)_{ESSFSM}}{\sigma(gg\ra h)_{SM}}  \frac{Br(h \rightarrow {\gamma\gamma})_{\rm ESSFSM}}{Br(h \rightarrow {\gamma\gamma})_{\rm SM}}.
\eeq
The production cross section of $h$ is same as in the SM. Then $\mu_{\gamma\gamma}$ can be written as
\beq
\mu_{\gamma\gamma} = \frac{\Gamma^{total}_{h,\rm SM}}{\Gamma^{total}_{h,\rm ESSFM}}, \,\,\,\, {\rm as}\,\,\Gamma_h^{total}/{M_h} \ra 0.
\label{Hdecay}
\eeq
As we do not have any extra charged particle, the decay width $\Gamma(h\rightarrow \gamma\gamma)$ is same as in the SM.
If the extra particles (scalar and fermions) have the mass less than half of the Higgs mass $M_h/2$, then the diphoton signal strength could be changed due to the invisible decay of the Higgs boson.
Using the global fit analysis~\cite{Global} that such an invisible branching ratio is less than $\sim 20 \%$, so the decay width in  eqn.~\ref{Hdecay} provides a suppression of about $\sim 80-100$ percent.
The present combined value of $\mu_{\gamma\gamma}$ by the ATLAS~ and CMS collaborations is $1.14^{+0.19}_{-0.18}$~\cite{Khachatryan:2016vau}. As the partial decay width of the Higgs to the heavy $Z_2$- and $Z_2'$-even neutrinos is zero, it can not alter $\mu_{\gamma\gamma}$. We also check that the mass region $M_{DM}<55$ GeV of the $Z_2'$-odd scalar and $Z_2$-odd neutrino DM along with $|\kappa| \gtrsim 0.004 $ and/or $|C_{h,3}| \gtrsim 0.2 $ are excluded at 2$\sigma$.

\subsection{Relic density and direct search limits}
The relic density of DM all alone restricts the allowed
parameter space. 
The parameter space of this model should also satisfy the
combined WMAP and Planck~\cite{Ade:2013zuv} imposed dark matter relic density constraint $\Omega_{\rm tot} h^2=0.1198\pm0.0026$. 
In our calculation, we use the {\tt micrOMEGAs}~\cite{Belanger:2014vza} to calculate the total relic density of the two DM particles. In this model, we find the correct relic density for the dark matter particles mass $M_{DM}<55$ GeV. However, these regions in the parameter spaces are ruled out from the invisible Higgs decay width and direct search data. In the following, we discuss the detailed constraints from direct detection of two-component dark matter particles.

The WIMPs, in particular, those that have non-vanishing weak
interactions with the SM and therefore can be tested. They are
actively being searched for in the direct detection experiments which look for their
nuclear scatterings in the deep underground detectors. 
If the DM scatters from atomic nucleus, then it leaves their signature in form of
a recoiled nucleus. However, no confirmed detection of the DM in the experimental laboratory has been made so far.
If a discovery is within the reach
of a near-future direct detection experiment then these experiments will be
able to constrain the WIMP properties such as its mass, DM-nucleus scattering cross
section and possibly spin.

As we have two-component DM, it is very difficult to distinguish these DM particles in the direct detection experiment.
The local number density
of the DM particles in the solar neighborhood that is important in determining the total number of event rate in the experiment.
It is not entirely straightforward
to determine which component dominates the event rate.
There have been only a few works regarding the direct detection of multi-component DM~\cite{Batell:2009vb,Profumo:2009tb,Herrero-Garcia:2017vrl}.
The signal rate generated from two-components DM in the detector is different than a single component DM and it completely depends on the DM masses and local densities in the solar neighborhood. The particle masses will determine their individual rates (see section 3.2 of the Ref.~\cite{Herrero-Garcia:2017vrl}) that can distinguish one or two-components DM if the DM particles have different masses.

\begin{figure}[h!]
 \begin{center}
  \subfigure[]{
 \includegraphics[width=2in,height=1.5in, angle=0]{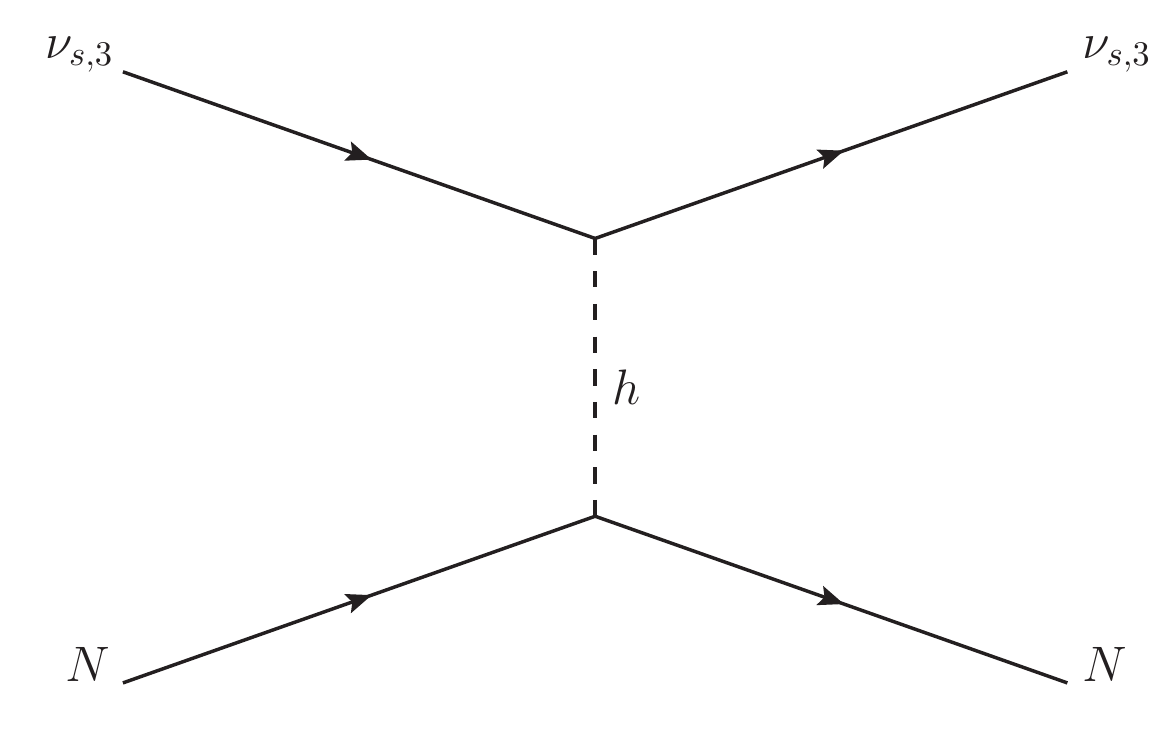}}
  \subfigure[]{
 \includegraphics[width=2in,height=1.5in, angle=0]{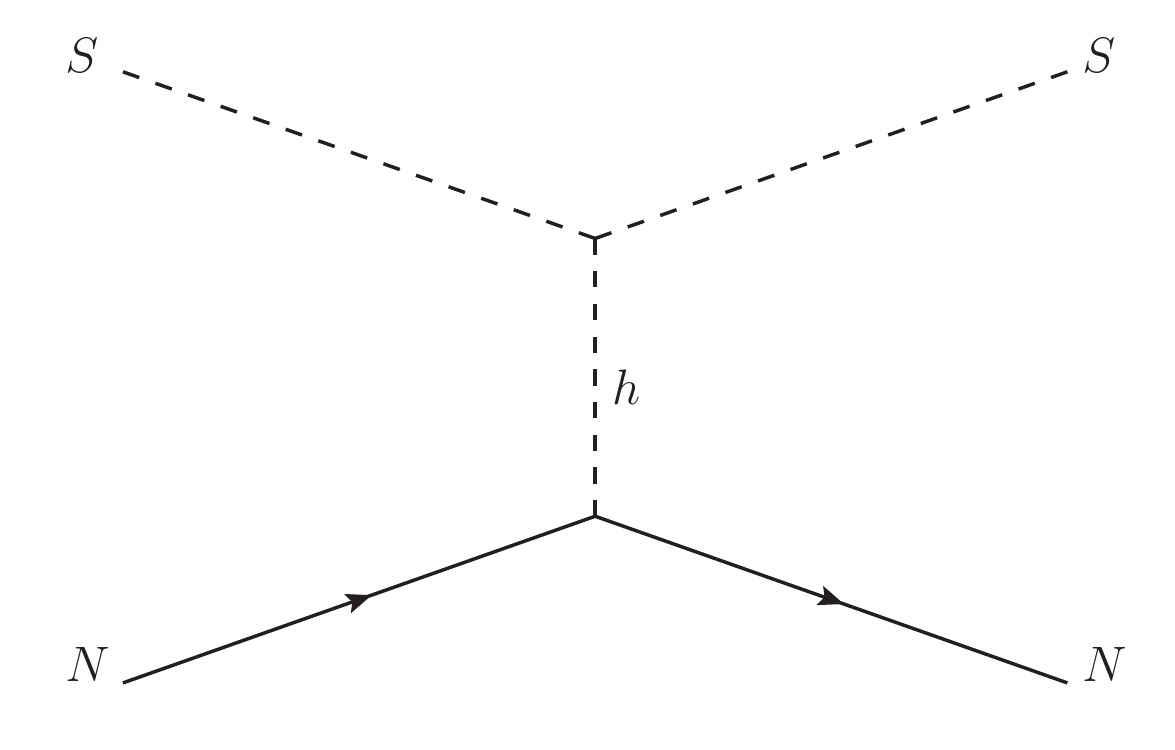}}
  \caption{\label{fig:Fendiag} \textit{\rm{ (a) Lowest order Feynman diagram for singlet neutrino-nucleus elastic scattering via the Higgs mediation. (b) A similar diagram for the singlet scalar-nucleus elastic scattering.}} } 
 \end{center}
 \end{figure}

Presently, non-observation of DM in the direct detection experiments such as XENON100~\cite{Aprile:2012nq}, LUX~\cite{Akerib:2013tjd, Akerib:2016vxi}, XENON1T~\cite{Aprile:2017iyp} set a limit on WIMP-nucleon scattering cross-section for a given DM masses. The most stringent bound is set by the XENON1T~\cite{Aprile:2017iyp} and LUX 2016~\cite{Akerib:2016vxi} exclusion data. The region above the green-line in Fig.~\ref{fig:crosMass} is excluded.
We translate the LUX exclusion data into some allowed or excluded zones in
the parameter spaces of our model comprising of $C_{h,3}$, $M_{{\nu_{s,3}}}$, $\kappa$, $M_S$ and $C_{S,3}$. In this model, the Feynman diagrams for the scattering of DM particles $\nu_{s,3},~S$ with the nuclei are shown in Fig.~\ref{fig:Fendiag}. In the limit $M_{DM} (M_{\nu_{s,3}},~M_S)\gg M_N$, the fermion-nucleon and scalar-nucleon scattering cross-sections are roughly given by
\beq
\sigma_{\nu_{s,3},N} = X_N \left(\frac{C_{h,3}}{\Lambda\, M_{\nu_{s,3}}}\right)^2~~~~~\text{and}~~~~~ \sigma_{S,N} = \frac{X_N}{2} \left(\frac{\kappa}{M_{S}}\right)^2\,. \label{directcs}
\eeq
where $X_N = \left(\frac{m_r m_N f}{ \sqrt{\pi} M_h^2}\right)^2$ and $f\approx0.3$ is the form factor of the nucleus. $m_r$ represents the reduced mass of the nucleus and the scattered DM particle.
%%%%%%%%%%%%%%%%%%%%%%%%%%%%%%%%%%%%%%%%%%%
%%%%%%%%%%%%%%%%%%%%%%%%%%%%%%%%%%%%%%%%%%%%

Using the eqns.~\ref{directcs}, we calculate the DM-nucleon cross-sections for the two dark matter components of different mass. The region in the parameter space for which DM-nucleon cross-section falls above the green-line in Fig.~\ref{fig:crosMass} is ruled out by the recent LUX-2016~\cite{Akerib:2016vxi} exclusion data. The region above the purple-line is ruled out by the recent XENON-2017~\cite{Aprile:2017iyp} data. 
\section{Numerical results}
\label{sec:result}
We explain the neutrino mass using the type-I seesaw mechanism. We show that our results are compatible with the various constraints such as the charged lepton flavor violating decay $\mu\rightarrow e \gamma$.
In addition the extra $Z_2$-odd fermion and the $Z_2^\prime$-odd scalar both can serve as viable DM particles producing the relic density in the right ballpark.
We show that the regions in the parameter space are consistent with the Planck/WMAP as well as LUX-2016, Xenon-2017 data.
In this study, we use {\tt FeynRules}~\cite{Alloul:2013bka} along with {\tt micrOMEGAs}~\cite{Belanger:2014vza} to compute the relic density of the DM candidates $\nu_{s,3}$ and $S$. We will discuss these in the following.

\subsection{Neutrino oscillation parameters}
%%%%%%%%%%%%%%%%%%%%%%%%%%%%%%%%%%%%%%%%%%%%%%%%%%%
\begin{table}[h!]
\begin{center}\scalebox{0.98}{
\begin{tabular}{|c||c|c|c|}
\hline
~~~~~~Parameters~~~~~~ &\multicolumn{3}{c|}{Benchmark Points for $Z_2$ and $Z_2^\prime$-even fermions}\\
\cline{2-4} 
 & ~~~BM-I ~~~& ~~~BM-II~~~& ~~~BM-III ~~~ \\
\hline
$M_{11}$~GeV&  $1.2\times10^{11}$&$7.1\times10^{5}$&$6.9\times10^{3}$\\

$C_{h,1}$~&  0.1&0.01&0\\

$M_{22}$~GeV&  $1.4\times10^{12}$&$2.36\times10^{5}$&$2.41\times10^{3}$\\

$C_{h,2}$~&  0.1&0.01&0\\

$M_{12}$~GeV&  0&0&0\\ 

$y_\nu$&      0.01 &$10^{-5}$&$10^{-6}$\\
  \hline
\hline
Outputs&\multicolumn{3}{c|}{ Corresponding Low-energy variables}\\
\hline
 $\Delta m^2_{21}/10^{-5} ~{\rm eV^{2}}$&         7.5001 &  7.2909 & 7.7197\\
  $ \Delta m^2_{31}/10^{-3} ~{\rm eV^{2}}$&           2.55234 & 2.63959  & 2.5312\\
         $ \theta_{12}$&            0.5883  &   0.5774   & 0.5720\\
         $ \theta_{23}$&            0.7953  &  0.7854   &   0.7803\\
          $\theta_{13}$&            0.1476  &  0.1473  & 0.1469\\
                $\delta_{PMNS}$~~ rad&            $10^{-5}$   &  $10^{-4}$  &  $10^{-3}$ \\
               $\alpha$~~ rad&            1.7 &  1.8  & 1.9\\
        $m_i$ ~eV&    ~~          0, 0.0087,  0.0505~~& ~~  0, 0.0085,0.0514~~  &  ~~ 0,   0.0088,   0.0505~~\\
$Br(\mu \rightarrow e~\gamma) $ & $ 3.0 \times 10^{-48} $ &   $1.9 \times 10^{-37}$ &  $   1.69 \times 10^{-33}$ \\
\hline
\end{tabular}}
\end{center}
\caption{ Three lists of benchmark points used in our analysis. Using these BPs, we have obtained the outputs for our model which are satisfying all the low energy constraints.}
\label{tab1}
\end{table}
%%%%%%%%%%%%%%%%%%%%%%%%%%%%%%%%%%%%%%%%%%%%%%%%%%%%%%%%

We obtain tiny neutrino mass through the type-I seesaw mechanism. We use the input parameters such as the new Yukawa couplings $ Y_{\nu,ij}$ ($i = 1,2,3$ and $j = 1,2$), dimensionless couplings $C_{h,1}$, $C_{h,2}$, $C_{h,12}$ and the mass terms $\overline{M}_{\nu_s,1}$, $\overline{M}_{\nu_s,2}$, and $\overline{M}_{\nu_s,12}$. In our calculation, we assume cut-off scale for the new physics is $\Lambda=10$ TeV. In order to explain successful leptogenesis~\cite{Covi:1996wh,Harvey:1990qw}, we need complex Yukawa coupling to have non-zero $CP$-violation.
The detailed discussion can be found in Ref.~\cite{Bambhaniya:2016rbb}. The presence of the extra Majorana neutrinos will allow for neutrinoless double $\beta$-decay~\cite{Mitra:2011qr}. In this work, we use the non-zero and real Yukawa couplings $Y_{\nu,12}= Y_{\nu,23} (\equiv y_\nu)$. Other Yukawa couplings are taken to zero.
We chose the values of the parameter $\overline{M}_{\nu_s,12}$ and $C_{h,12}$ such that the off-diagonal components of the heavy mass matrix ${M}_{\nu_s}$ become zero (see eqn.~\ref{MasMatrixs}). We consider three heavy neutrino masses $\mathcal{O}(10^{11})$ GeV, $\mathcal{O}(10^{5})$ GeV and $\mathcal{O}(10^{3})$ GeV and corresponding three different Yukawa couplings $y_\nu$ to obtain tiny the neutrino masses. We present these benchmark points and the corresponding low energy variables in Table~\ref{tab1}. These variables are consistent with the experimental data. As the cut-off scale for the new physics $\Lambda$ is very large, the dimensionless couplings $C_{h}$ (within perturbative limit) could not alter the neutrino mass considerably.

\subsection{New regions in the DM parameter space}
We have seen that the region in the parameter space $M_{DM}< \mathcal{O}(500)$ GeV of a single Higgs portal WIMP DM  particle are ruled out by the recent LUX experiment.
Hence, it becomes important to show that these regions in the parameter space are still alive in the ESSFSM. In Table~\ref{tab2}, we present five such benchmark points for this model which are producing right relic density and allowed by the recent non-observation of DM-nucleon scattering in the LUX experiment. The DM mass regions below the half of the Higgs mass are also consistent with the Higgs invisible decay width~\cite{Global}. 
If the mass difference between the fermionic and scalar DM
particles are very large, then it is expected that lighter one will dominate over the heavier one in contributing to the relic density. 
%%%%%%%%%%%%%%%%%%%%%%%%%%%%%%%%%%%%%%%%%%%%
\begin{table}[h!]
\begin{center}
\scalebox{0.98}{
\begin{tabular}{|c|c|c|c|c|c|c|c|c|c|c|}
\hline
Bench-&\multicolumn{5}{c|}{Parameters} & Relic& \multicolumn{2}{c|}{Percentage of DM} &\multicolumn{2}{c|}{DM-N cross-section in [zb]}\\
\cline{2-6}\cline{8-9}\cline{10-11}
mark&&&&&&density&&&& \\
Points & $M_{\nu_{s,3}}$ GeV & $C_{h,3}$& $M_{S}$~GeV & $\kappa$ &$~C_{S,3}~$ & $\Omega h^2$ & ~Fermion~ &  Scalar & ~~~Fermion~~~ & Scalar\\
\hline
&&&&&&&&&& \\
BP-I  &260& 0.05& 59& 0.0015&0.1& 0.1271& 39.12& 61.88& 0.33& 0.0054\\
&&&&&&&&&& \\
BP-II  &130&0.01 & 60 &0.001 & 0.1& 0.1263 & 40.68& 59.32 &0.013& 0.0023\\
&&&&&&&&&& \\
BP-III  &86&-0.01 & 59.8 &0.0012 & 0.1& 0.1129 & 53.10& 46.90 &0.013&0.0033\\
&&&&&&&&&& \\
BP-IV  &62& -0.01& 60.9& 0.0016&0.1& 0.1156& 69.76& 30.24& 0.013& 0.0058\\
&&&&&&&&&& \\
BP-V  &59& 0.02& 250& 0.0025&0.1& 0.1201& 99.6& 0.4& 0.053& 0.0008\\
&&&&&&&&&& \\
\hline
\end{tabular}}
\end{center}
\caption{Lists of Benchmark points used in our analysis. Using these BPs we obtain the relic density in the right ballpark allowed by LUX-2016 direct detection data.}
\label{tab2}
\end{table}
%%%%%%%%%%%%%%%%%%%%%%%%%%%%%%
For $M_{\nu_{s,3}} \simeq M_S$ and tiny interaction coupling $C_{S,3}$, the contribution of these DM particles are nearly equal into the total relic density. Whereas the interaction coupling $C_{S,3}\sim \mathbf{O}(1)$ and a huge mass difference in the DM particles with particular Higgs portal couplings $\kappa$ and $C_{h,3}$ can produce equal relic density in the Universe. For example, see the benchmark points I$-$IV. The lighter DM mass near 60 GeV will always dominate over the heavier one because the contribution of self-annihilation processes $DM, DM \rightarrow b\bar{b}$ into the relic density are larger than the other processes. The other processes can dominate over the $DM, DM \rightarrow b\bar{b}$ process for the choice of the large Higgs portal coupling $\kappa$. In this case, the relic density and the direct detection data restrict such a choices of $\kappa$.

\begin{figure}[h!]
 \begin{center}
  \subfigure[]{
 \includegraphics[width=2.7in,height=2.7in, angle=0]{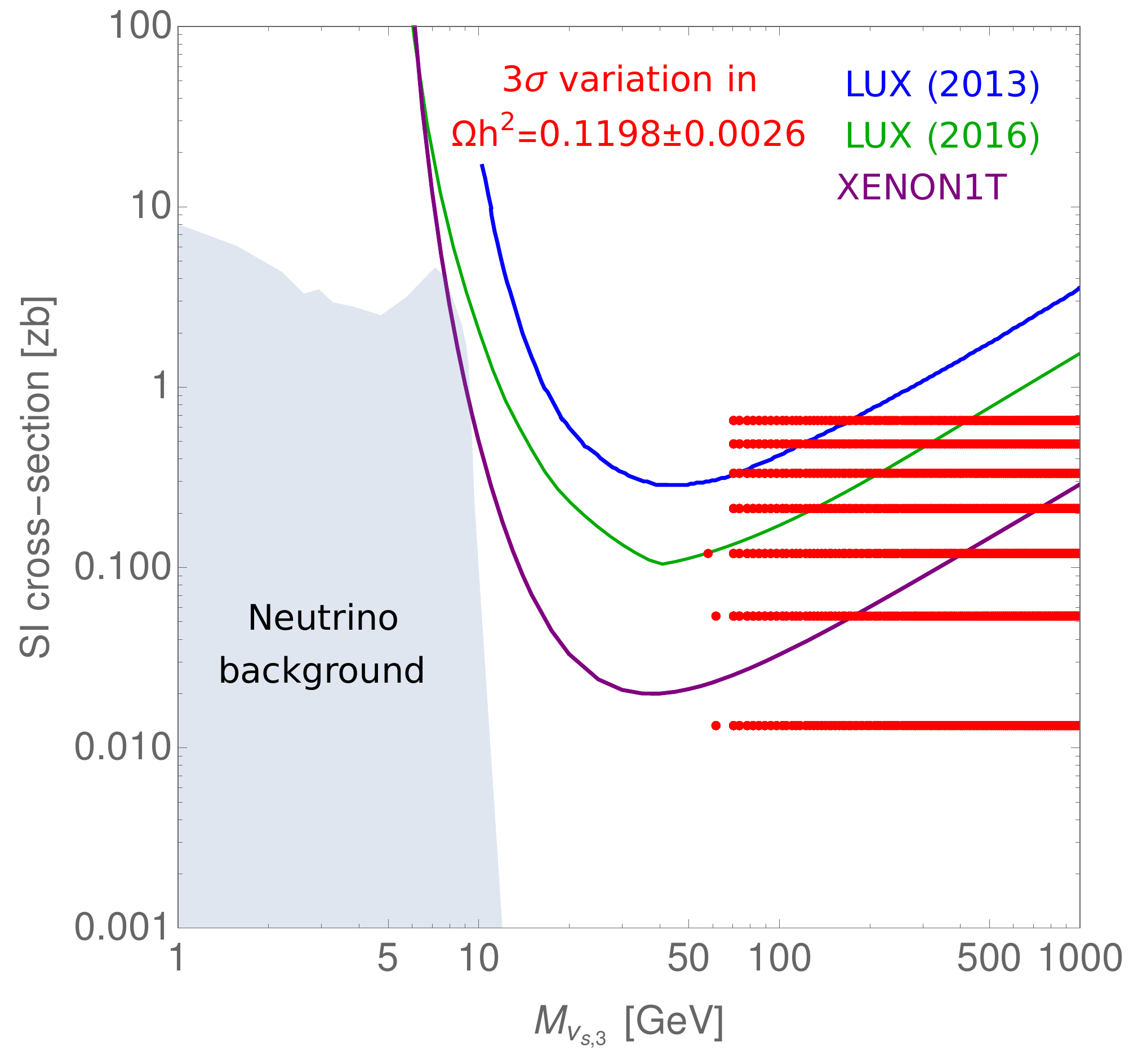}}
  \subfigure[]{
 \includegraphics[width=2.7in,height=2.7in, angle=0]{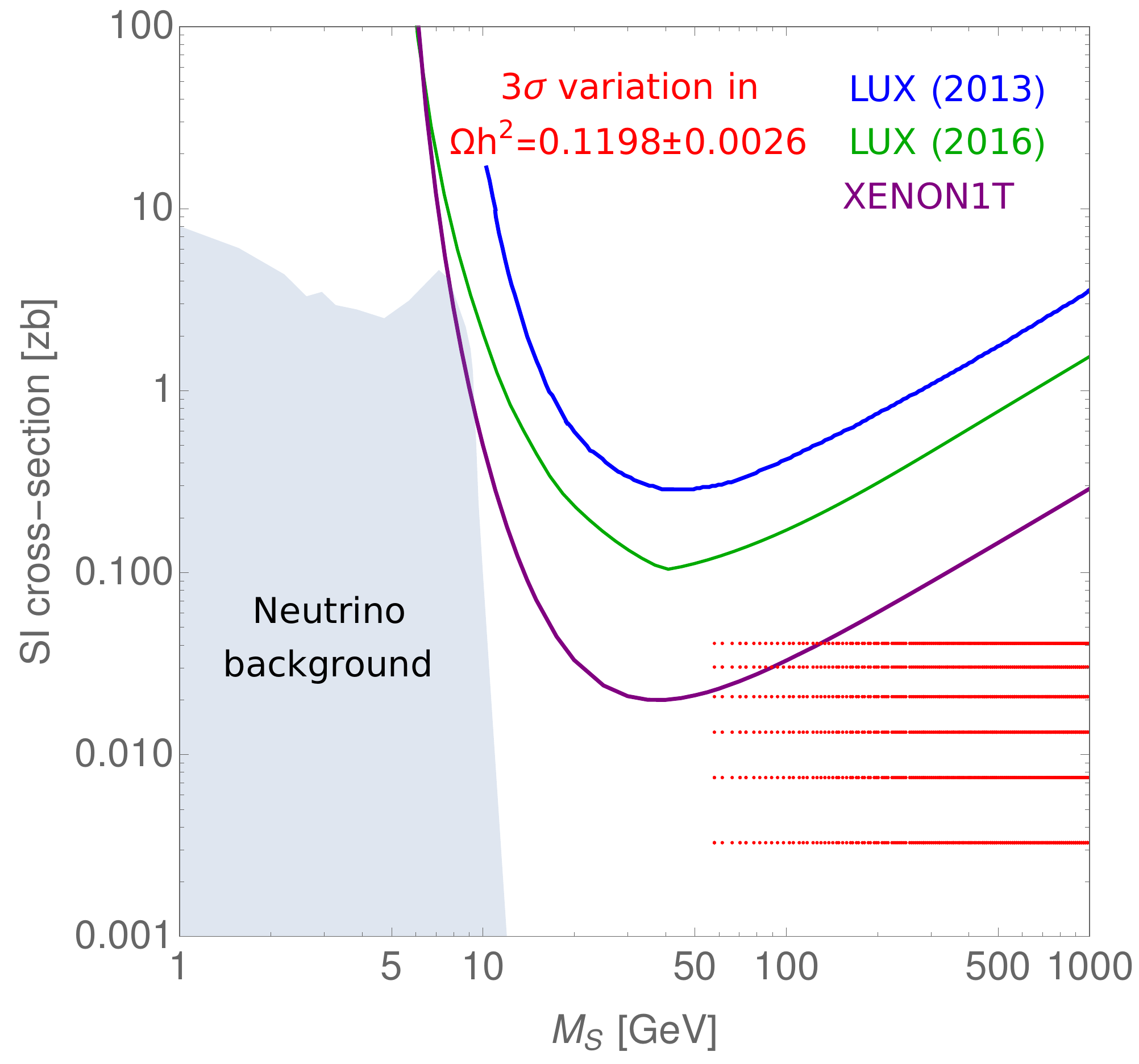}}
  \caption{\label{fig:crosMass} \textit{\rm{WIMP-nucleon cross section vs. DM mass keeping $C_{S,3}=0.1$. The gray region indicates the neutrino background. Note that each plot contains only $10^7$ data (red) points. Larger data points can fill the gap between the red bands.}} } 
 \end{center}
 \end{figure}

In order to find the favored regions in the parameter space which
satisfies DM relic density constraints and the recent LUX direct detection data, in Fig.~\ref{fig:crosMass} we
present two contour plots of relic density $\Omega h^2$ in the DM nucleon cross-section $vs$ mass plane.
The red-points consistent with the relic density $\Omega h^2=0.1198\pm0.0026$ within 3$\sigma$.
We vary $C_{h,3}$ from $-$0.7 to
0.7 and $\kappa$ from 0 to 0.75 to obtain Fig.~\ref{fig:crosMass}. We also fix the coupling $C_{S,3}=0.1$ in these plots. 
%%%%%%%%%%%%%%%%%%%%%%%%%%%%%%%%%%%%%%%%%%%%%%%%%%%%%%%
%%%%%%%%%%%%%%%%%%%%%%%%%%%%%%%%%%%%%%%%%%%%%%

\begin{figure}[h!]
 \begin{center}
\subfigure[]{
 \includegraphics[width=2.7in,height=2.7in, angle=0]{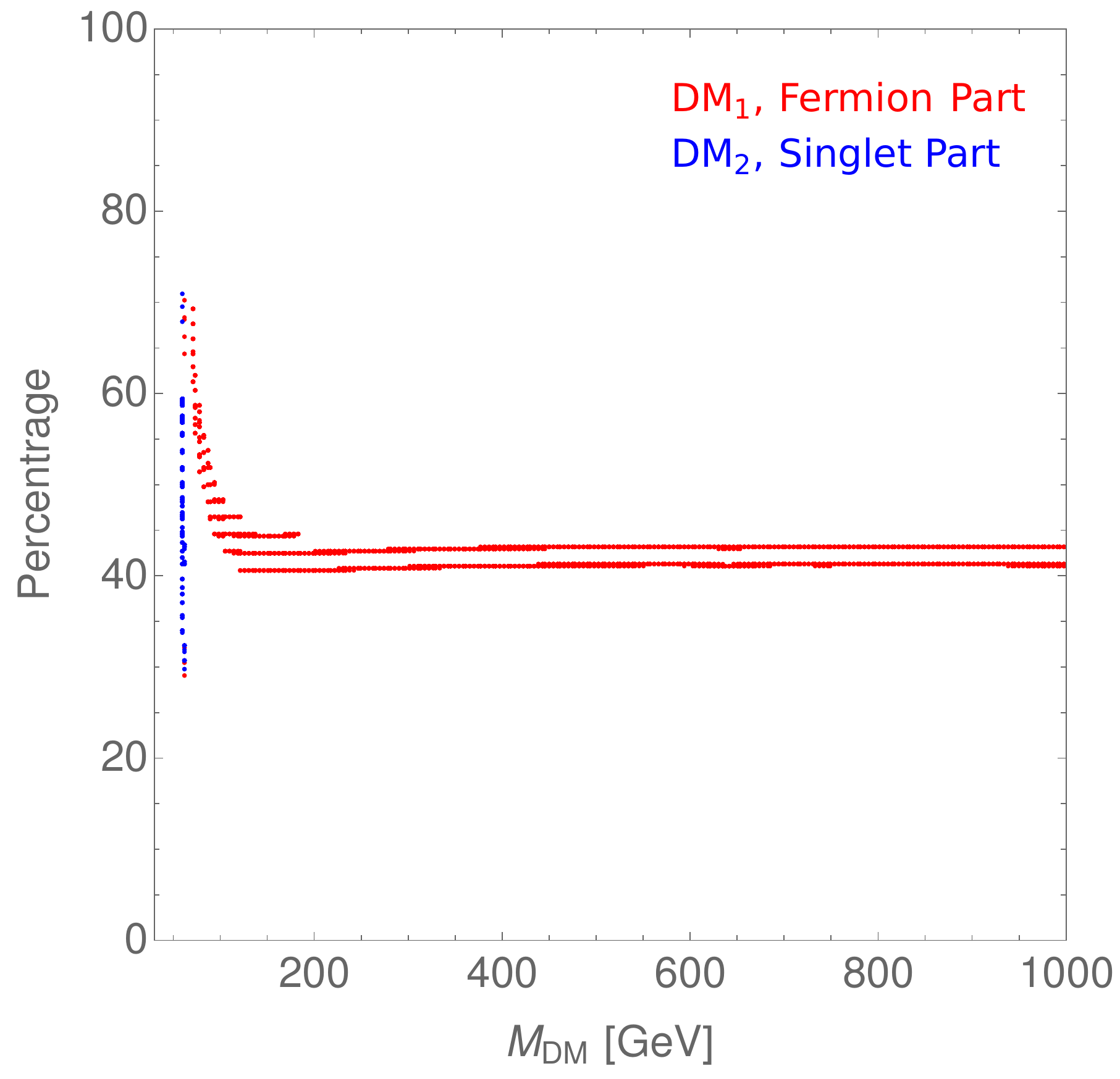}}
 \subfigure[]{
 \includegraphics[width=2.7in,height=2.63in, angle=0]{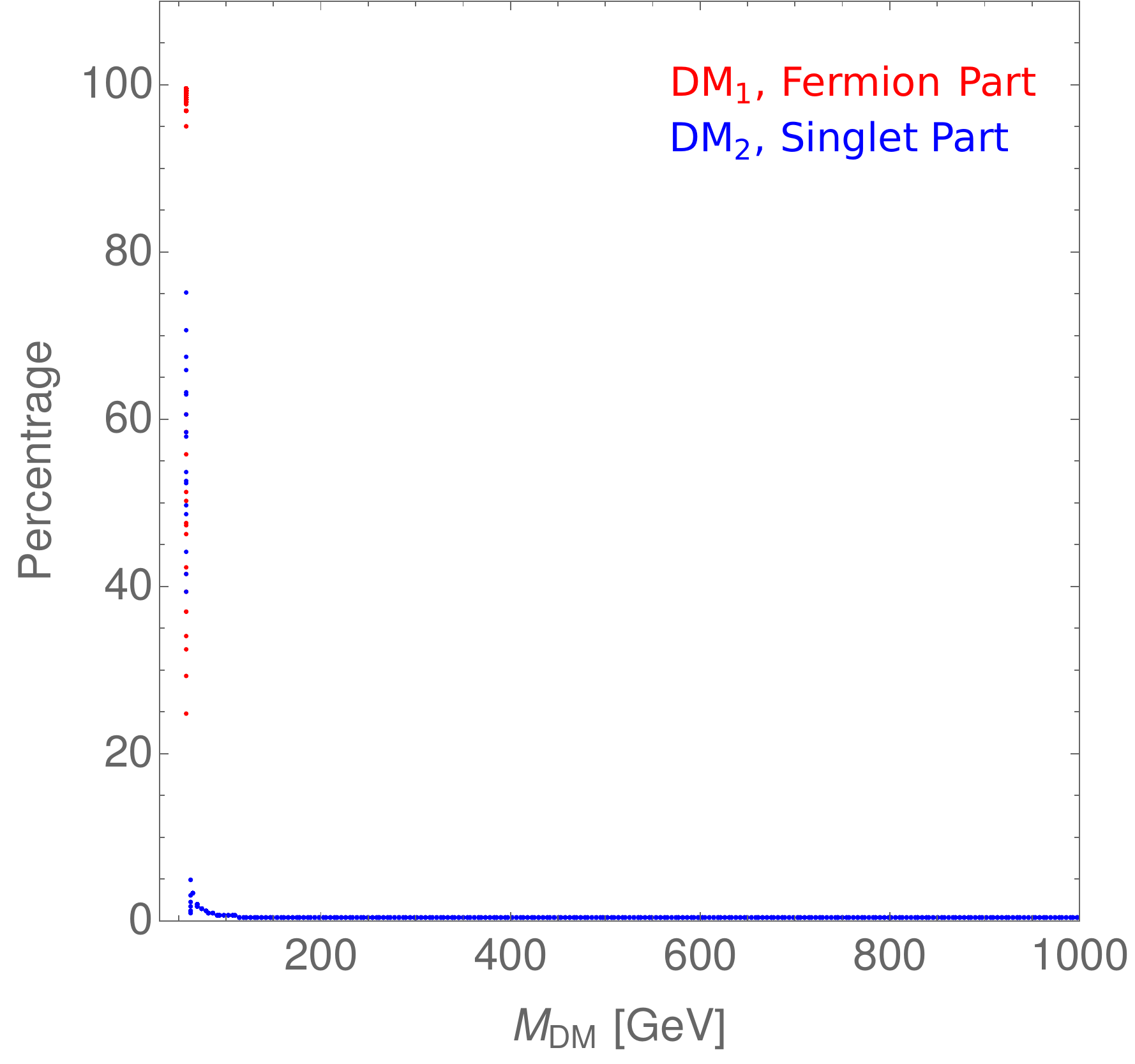}}
  \caption{\label{fig:percentage} \textit{\rm{Percentage of DM contributing to the total relic density. Red points correspond to the fermionic DM contributions whereas blue points for the singlet scalar.}} } 
 \end{center}
 \end{figure}

In Fig.~\ref{fig:crosMass}(a), we vary the scalar DM mass between 55 GeV and 65 GeV, the fermion DM mass between 40 and 1000 GeV. In Fig~\ref{fig:crosMass}(b), we take the variation of the fermionic DM mass between 55 GeV and 65 GeV and the scalar mass between 40 and 1000 GeV. We find that a large region in the parameter space satisfies the bound on WIMP-nucleon cross section as imposed by the recent LUX-2016 and Xenon-2017 experimental data.  
We find that the scalar DM mass $M_S\sim 60$ Gev provides the dominant contributions in the relic density. The contribution decreases with $M_S$. However, we need this scalar part to achieve the relic density as observed by the WMAP/Planck.
In the second case, the fermionic contribution remains same ($\sim50~\%$) in the region $55\lesssim M_S \lesssim 65$ GeV. We show these variations of the DM contribution in Fig.~\ref{fig:percentage} with the DM mass. The red points indicate the fermionic contribution whereas blue points stand for the scalar contribution to the correct relic density ($\Omega h^2 = 0.1198 \pm 0.0026$) within 3$\sigma$.
%%%%%%%%%%%%%%%%%%%%%%%%%%%%%%%%%%%%%%%%%
%%%%%%%%%%%%%%%%%%%%%%%%%%%%%%%%%%%%%%%%%

\section{Conclusion}
\label{sec:conclusion}
In this paper, using two right-handed singlet fermions, we have explained the neutrino mass through the type-I seesaw mechanism. 
We have chosen three representatives ``benchmark points" of three different Majorana mass parameter spaces ($\sim 10^3,~10^{6}$ and $10^{12}$) and particular structure of the Yukawa couplings matrix, i.e., the Dirac mass matrix to explain the neutrino mass-squared differences as observed by the neutrino experiments. We have also calculated the PMNS mixing angles and the other low-energy variables, e.g., non-unitarity constraints on the PMNS matrix, LFV constraints from $\mu\rightarrow e\gamma$, etc. The combinations of the new Yukawa couplings and the heavy neutrino mass are satisfied the neutrino mass and mixing angles constraints~\cite{Halprin:1976mr}.

In the presence of $Z_2$ and $Z_2^\prime$ symmetries, we have also analysed the two-component Higgs portal self-annihilating dark matter particles. 
The regions of mass $65-550$ GeV of a Higgs portal fermionic or scalar dark matter models are excluded by the recent LUX experiment. In this model, we have shown that the regions of the parameter space with two-component dark matter particles are still allowed from direct search experiment and the WMAP/Planck data. For different fermionic Higgs portal coupling $C_{h,3}$ and fermion dark matter mass, we have obtained viable scalar dark matter mass between $50$ GeV and $\sim 300$ TeV. We have also obtained the similar region of fermionic dark matter mass for different scalar Higgs portal coupling $\kappa$. The unitary bounds are violated the dark matter mass above 300 TeV~\cite{Griest:1989wd}.
Here, we do not intend to show that all the parameter spaces satisfy the experimental results. Rather, in the framework of our model, we have wanted to use the advantage of a two-component dark matter which has a large region of parameter spaces satisfying the constraint of various dark matter experiments.

The model ESSFSM is considered here to present the minimal seesaw mechanism and two-component dark matters in terms of particles content. This model can explain the observed tiny neutrino mass-squared differences and the mixing angles in oscillation experiments. The regions in the parameter space are also consistent the relic density of dark matter observed by the Planck, WMAP experiments and the recent null-results of the WIMPs dark matter from the direct search LUX-2016 and XENON-2017 experiments.

\vskip 20pt
%\section{Acknowledgements}
\noindent{\bf Acknowledgements}\\
We would like to thank Subhendu Rakshit and Vishnudath K. N. for useful discussions.

\end{document}